\documentclass[conference]{IEEEtran}
\usepackage{xcolor}

\usepackage{subfigure}
\usepackage{tikz,pgfplots}
\usepackage{graphics}
\usepackage{amsmath}

\usepackage{xcolor}

\definecolor{red}  {rgb}{0.9,0.0,0.0}
\definecolor{green}{rgb}{0.0,0.7,0.0}
\definecolor{blue} {rgb}{0.0,0.0,0.9}

\usepackage{amsfonts}
\usepackage{amsmath,mathtools,commath,nicefrac}

\usepackage{graphicx,float}
\usepackage{wrapfig}

\usepackage{tabularx}
\usepackage{booktabs}

\usepackage{subfigure}

\usepackage{amsthm}

\usepackage{tikz}
\usepackage{gnuplot-lua-tikz}
\usetikzlibrary{shapes,decorations,shadows,positioning,chains,fit,shapes,calc,matrix,backgrounds,arrows}

\usetikzlibrary{quotes}
\usetikzlibrary{decorations.pathreplacing,positioning, arrows.meta}

\usepackage{siunitx}

\usepackage{tikz}
\usetikzlibrary{shapes,arrows.meta,calc,decorations,arrows}
\usetikzlibrary{chains}
\usepackage{pgfplots}
\pgfplotsset{compat=1.18} 

\usetikzlibrary {patterns,patterns.meta}

\tikzstyle{nstyle}=[draw,circle, minimum size=12,fill=white,inner sep=0.2pt]
\tikzstyle{estyle}=[draw,line width=1 pt]
\tikzstyle{estyle-emph}=[draw,line width=2 pt,blue]
\tikzstyle{graphstyle}=[fill=gray!20]
\tikzstyle{smallnode}=[draw,circle, minimum size=1, inner sep=1]
\tikzstyle{snstyle}=[draw,circle, minimum size=2,fill=white,inner sep=0.1pt]
\tikzstyle{rootstyle}=[snstyle]
\tikzstyle{sestyle}=[draw]

\usepackage{pgfplots}
\usepackage{xifthen}

\usetikzlibrary{svg.path}

\usepackage{comment}
\newtheorem{property}{Property}%
\newtheorem{cnstr}{\textbf{Construction}}

\theoremstyle{definition} 

\newcommand{\withrio}[1]{}
\newcommand{\extended}[1]{}
\newcommand{\added}[1]{{\color{black}#1}}

\begin{document}

\title{Network Topologies  for QKD Networks}


\author{
\IEEEauthorblockN{Ori Rottenstreich\\}
\IEEEauthorblockA{Technion}
\and
\IEEEauthorblockN{Ran Hasson Ruso\\}
\IEEEauthorblockA{NVIDIA}
\and
\IEEEauthorblockN{Eliahu Cohen\\}
\IEEEauthorblockA{Bar-Ilan University}
}

\pagestyle{empty}
\maketitle

\begin{abstract}
Quantum key distribution (QKD) is a method for distributing cryptographic keys between remote endpoints, enjoying security based on quantum physics. This paper makes a first step towards studying topologies for QKD networks. QKD networks imply several required characteristics for their reliability and efficiency. We express such properties in terms of graph structure. For the comparison of potential graphs, we describe cost functions that allow us to identify families of graphs that are reliable and efficient. To make the discussion realistic, we summarize representative field-reported QKD performance numbers and illustrate the sensitivity of secret-key rate (SKR) to a few dB of additional loss. Last, we present 
methods to construct large efficient graphs through connecting smaller graphs. 
\end{abstract}


\section{Introduction}
\thispagestyle{empty}
Quantum key distribution (QKD) is a method for distributing cryptographic keys between two entities, enjoying security based on quantum physics. QKD is one of the most mature technologies among emerging technologies that leverage quantum capabilities for outperforming conventional methods in terms of functionality and performance~\cite{pirandola2020advances}. A physical QKD system is a communication system that exploits quantum physics to facilitate a distribution of quantum correlations between parties that can lead to joint computation of a secret key. QKD~\cite{mehic2020quantum, ZhuYZNZ23}  relies on three main foundational principles known as (i) Quantization and indivisibility of quanta; (ii) Quantum uncertainty; (iii) No-cloning theorem. 

Several recent studies aim to examine the integration of QKD in datacenters to enhance  security~\cite{jain2023quantum, ZhuYZNZ23}. 
Datacenter networks are emerging to satisfy the computation, communication and caching requirements of growing services. Many of these services also have high security requirements. These can include, among others, hospitals, smart cities, banking and military applications. The security of traditional key-distribution techniques can be compromised with the developments in quantum computing.  

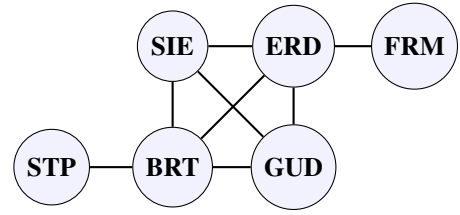
\begin{figure}[t!]
    \centering
        \begin{tikzpicture}[
                        scale=0.80, 
    node/.style={
        circle, 
        draw=black, 
        fill=blue!5, 
        minimum size=0.18cm, 
        font=\bfseries
    },
    edge/.style={
        thick, 
        black
    },
    cnode/.style={
        circle, 
        draw=black, 
        pattern={Dots[distance=1.7pt, radius=0.45pt]},
        pattern color=red!60,
        minimum size=0.18cm, 
        font=\bfseries
    },
]

    \node[node] (A) at (0,0) {BRT};

    \node[node] (B) at (0,2) {SIE};
    \node[node] (C) at (2,2)  {ERD};
    \node[node] (D) at (2,0){GUD};
    \node[node] (E) at (-2,0){STP};
    \node[node] (F) at (4,2){FRM};
    
    \draw[edge] (A) -- (B);
    \draw[edge] (A) -- (C);
    \draw[edge] (A) -- (E);
 \draw[edge] (B) -- (C);
    \draw[edge] (B) -- (D);
        \draw[edge] (C) -- (F);
        \draw[edge] (C) -- (D);
        \draw[edge] (A) -- (D);
\end{tikzpicture}
\caption{Example of a real QKD Topology of the SECOQC QKD network with six nodes and eight edges. Node's names are mentioned as their original names. Source: Mehic et al.~\cite{mehic2020quantum}.}
    \label{fig:real_example}
\end{figure}

While traditional datacenters often have 3C requirements such as communication, computation and caching, with QKD they also include a fourth fundamental requirement of cryptographic capabilities. Jointly they are often known as 4C. The design of traditional datacenters has been studied over the years, in multiple dimensions, such as their physical structure (topology)~\cite{Al-FaresLV08, DragonflyPlus, hyperx}, resource allocation  (e.g., mapping of tenants)~\cite{tree_allocation} and routing for communication among hosts of the same application (path selection)~\cite{ZahidGBJS17}. Fig.~\ref{fig:real_example} illustrates the Vienna SECOQC trusted-relay QKD network, one of the earliest multi-node QKD network demonstrations \cite{peev2009secoqc} (see also \cite{mehic2020quantum}). 
The FP6 Project SECOQC (Secure Communication
based on Quantum Cryptography)  involved 41 research and industrial partners from 11
European Union countries, Russia, and Switzerland in 2004. 


The traditional approaches in the three dimensions might not optimize QKD networks due to some special properties: 

\emph{(i)} (Robustness) QKD networks  should be able to handle node and link failures quickly. This should be taken into account, e.g, in selecting multiple  paths which can be disjoint with regards to their intermediate nodes or links. 

\emph{(ii)} (Limited path length) \added{In practice, each quantum link is loss-limited: fibre attenuation (added optical loss, in dB) and component insertion loss directly reduce the secret-key rate (SKR) and can drive it to zero. End-to-end key delivery across multiple hops is typically achieved by trusted relays and key management~\cite{dervisevic2026kms}; therefore, hop count mainly increases the trust assumptions and consumes link key capacity, rather than ``accumulating quantum noise'' end-to-end~\cite{mehic2020quantum}.}

\emph{(iii)} (Routing criteria) Beyond path length, additional criteria can affect path selection, such as the level of trust in nodes along the path. Including vulnerable nodes along the path can harm security.

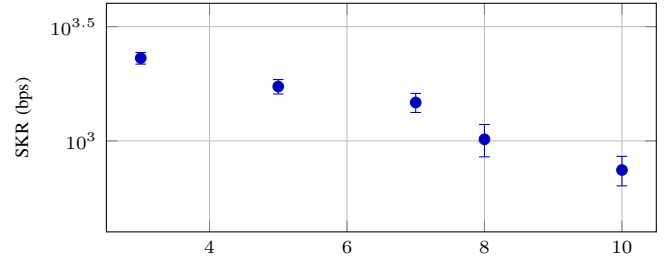
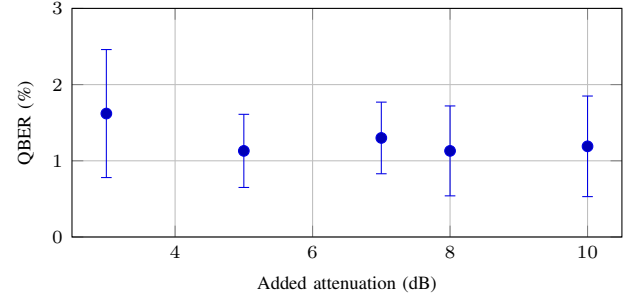
\begin{figure}[t!]
\centering
 \subfigure[SKR vs. added attenuation.]{
\centering
\begin{tikzpicture}
\begin{axis}[
    width=\columnwidth,
    height=0.52\columnwidth,
    xlabel={Added attenuation (dB)},
    ylabel={SKR (bps)},
    ymode=log,
    grid=both,
    xmin=2.5, xmax=10.5,
    ymin=400, ymax=4000,
    tick label style={font=\scriptsize},
    label style={font=\scriptsize},
]
\addplot+[
    only marks,
    error bars/.cd,
      y dir=both, y explicit,
] coordinates {
    (3,2303) +- (0,135)
    (5,1730) +- (0,126)
    (7,1473) +- (0,141)
    (8,1016) +- (0,164)
    (10,746) +- (0,110)
};
\end{axis}
\end{tikzpicture}
}
\hfill
 \subfigure[QBER vs. added attenuation.]{
\centering
\begin{tikzpicture}
\begin{axis}[
    width=\columnwidth,
    height=0.52\columnwidth,
    xlabel={Added attenuation (dB)},
    ylabel={QBER (\%)},
    grid=both,
    xmin=2.5, xmax=10.5,
    ymin=0, ymax=3.0,
    tick label style={font=\scriptsize},
    label style={font=\scriptsize},
]
\addplot+[
    only marks,
    error bars/.cd,
      y dir=both, y explicit,
] coordinates {
    (3,1.62) +- (0,0.84)
    (5,1.13) +- (0,0.48)
    (7,1.30) +- (0,0.47)
    (8,1.13) +- (0,0.59)
    (10,1.19) +- (0,0.66)
};
\end{axis}
\end{tikzpicture}
}
\caption{Sensitivity of QKD key parameters to added attenuation in a commercial datacenter interconnect field trial (values from Table~3 in~\cite{qiu2024dci}).}
\label{fig:attenuation_sensitivity}
\end{figure}

\added{\emph{(iv)} (Loss sensitivity and hardware insertion loss) In a datacenter, a ``logical'' edge may traverse patch panels, wavelength division multiplexing (WDM) components, or optical switches; even a few dB of extra loss can materially reduce SKR, and beyond a loss budget the SKR can drop to zero. This motivates making topological comparison explicitly loss-aware (see Fig.~\ref{fig:attenuation_sensitivity}).}

Note that below we focus on trusted-relay (hop-by-hop) QKD networks. Quantum repeaters are beyond the scope of this work.

{\bf QKD terminology for networking readers.} 

\emph{(i)} We use secret-key rate (SKR) as the link-level key-generation capacity: after the quantum transmission, reconciliation, and privacy-amplification steps, it is the rate at which two adjacent QKD devices obtain shared secret bits. 

\emph{(ii)} Quantum bit error rate (QBER) is the fraction of sifted key bits on which the two endpoints disagree. It reflects noise, device imperfections, and other errors, and it directly affects how much secret key can be extracted. We focus on trusted-relay QKD networks, where an end-to-end key over multiple hops is supported by hop-by-hop QKD keys and intermediate relays must be trusted. Thus, path length is cruical not because a quantum signal traverses many hops end-to-end, but because each hop consumes link key capacity and each relay increases the trust surface. 

\emph{(iii)} Finally, each physical QKD link has a finite optical-loss budget, including fiber loss and insertion loss from connectors, patch panels, WDM components, or optical switches. Since attenuation is measured in dB, a few additional dB can substantially reduce photon arrival probability and therefore SKR. Beyond a platform-dependent threshold, the net key rate can become effectively zero.

{\bf From graph metrics to QKD practice.} We first use unweighted graph metrics as protocol-agnostic proxies for (i) robustness (critical nodes, disjoint paths) and (ii) key-transport overhead (mean hop distance) in trusted-relay QKD networks. 
We then make the same comparison loss-aware by assigning each edge a loss $L_e$ and a key-generation capacity $K_e \approx \mathrm{SKR}(L_e)$ (Sec.~\ref{section_basic_graph_properties}--\ref{section_cost_functions}), linking $\bar d$ and $|E|$ to delivered key rate.
Disjoint paths can also enable multipath key splitting to reduce reliance on any single relay (Sec.~\ref{section_cost_functions}).

\textbf{Field numbers and loss sensitivity.}
Table~\ref{tab:field_qkd_stats} summarizes indicative (and highly variable) field-reported QKD performance points, as collected in~\cite{qiu2024dci}. Fig.~\ref{fig:attenuation_sensitivity} reproduces a datacenter interconnect attenuation test from~\cite{qiu2024dci}, showing how SKR degrades as added attenuation increases. These numbers are not meant as universal benchmarks; rather, they motivate why a topology ``cost function'' should include edge loss budgets (fibre length + insertion loss) and not only hop count.

\begin{table}[t!]
\centering
\scriptsize
\begin{tabular}{@{}p{0.42\columnwidth}p{0.22\columnwidth}p{0.28\columnwidth}@{}}
\toprule
Deployment (examples) & Distance / loss & Reported performance \\
\midrule
Vienna SECOQC (2008) & 33 km, 7.5 dB & SKR 3.1 kbps, QBER 2.6\% \\
Switzerland (150 km link) & 150 km, 43 dB & SKR 2.5 bps, QBER 5\% (COW) \\
China metro (2012) & 85.1 km, 18.4 dB & SKR 0.77 kbps, QBER 5.26\% (decoy BB84) \\
Cambridge UK network & 10.6 km, 3.9 dB & SKR 2580 kbps, QBER $<2.5$\% \\
Commercial datacenter interconnect & \textit{(see~\cite{qiu2024dci})} & Avg. SKR 2.392 kbps, QBER $\approx1.9$\% \\
\bottomrule
\end{tabular}
\caption{Indicative field-reported QKD performance highlights (summarised in~\cite{qiu2024dci} and references therein).}
\label{tab:field_qkd_stats}
\end{table}

We focus on the following research avenues: 
\begin{itemize}
    \item Developing cost metrics for networks optimized for QKD in terms of their overhead and robustness. 
    \item Designing new topologies for QKD networks which can 
    have a good tradeoff between robustness and the implied overhead.   
  \item Discussion on the extension of the results beyond graph structure. 
\end{itemize}

Table~\ref{table_notations} summarizes the main notations of the paper.

\begin{table}[t]
\centering
\caption{Summary of Main Notations.}
\label{table_notations}
\begin{tabular}{cl}
\toprule
\textbf{Notation} & \textbf{Meaning} \\
\midrule
$n$ & Number of participating  nodes \\
$O$ & Operational cost for nodes and edges of a graph\\
$R$ & Reliability factor of a graph\\
$C$ & Connectivity factor of a graph\\
$\Phi(G), \Psi(G)$ & Cost functions for a graph\\ 
$\alpha, \beta, \gamma$ & Parameters of graph cost functions\\
$D$, $\bar{d}$ & maximal distance (diameter) and mean distance among nodes\\
\bottomrule
\end{tabular}
\end{table}

\begin{table*}[t!]
\centering
\scriptsize
\resizebox{0.98\textwidth}{!}{%
\begin{tabular}{c|lccccccc}
\hline
Family index & Graph Example  & \# Edges & Mean degree & Max degree & Mean Distance & Diameter & Non-critical & Pairs with  \\
& (list of edges)  & & & & & &  Nodes & Disjoint Paths \\
\hline
      1     &                                      (A,B), (A,C), (A,D), (A,E)     &    4   &    1.6   &     4   &     1.6   &      2    &          4  &               0  \\
      2   &                                        (A,B), (A,D), (A,E), (B,C)      &   4  &     1.6    &    3   &     1.8   &      3     &         3 &                0\\
      3    &                                       (A,C), (A,E), (B,C), (B,D)     &   4  &     1.6    &    2   &     2.0   &      4    &          2  &               0 \\
      4     &                               (A,B), (A,C), (A,D), (A,E), (B,C)      &  5  &     2.0    &    4   &     1.5   &      2    &          4   &              3 \\
      5      &                              (A,B), (A,C), (A,E), (B,C), (B,D)      &  5  &     2.0    &    3   &     1.6   &      3    &          3   &              3 \\
      6       &                             (A,C), (A,D), (A,E), (B,C), (B,D)     &  5  &     2.0    &    3   &     1.6   &      3    &          4   &              6 \\
      7        &                            (A,B), (A,E), (B,C), (B,D), (C,D)      &  5  &     2.0    &    3   &     1.7   &      3    &          3   &              3 \\
      8     &                               (A,D), (A,E), (B,C), (B,E), (C,D)     &  5  &     2.0    &    2   &     1.5   &      2    &          5   &             10 \\
      9    &                         (A,B), (A,C), (A,D), (A,E), (B,C), (B,D)       &  6  &     2.4    &    4   &     1.4   &      2    &          4   &              6 \\
     10    &                         (A,C), (A,D), (A,E), (B,C), (B,D), (B,E)      &  6  &     2.4    &    3   &     1.4   &      2    &          5   &             10 \\
     11    &                         (A,B), (A,C), (A,E), (B,C), (B,D), (C,D)     &  6  &     2.4    &    3   &     1.5   &      3     &         4   &              6 \\
     12    &                         (A,B), (A,C), (A,D), (A,E), (B,E), (C,D)      &  6  &     2.4    &    4   &     1.4   &      2    &          4   &              6 \\
     13    &                         (A,B), (A,D), (A,E), (B,C), (B,E), (C,D)    &   6  &     2.4    &    3   &     1.4   &      2    &          5   &             10 \\
     14   &                   (A,B), (A,C), (A,D), (A,E), (B,C), (B,D), (B,E)    &   7  &     2.8    &    4   &     1.3   &      2    &          5   &             10 \\
     15   &                   (A,B), (A,C), (A,D), (A,E), (B,C), (B,D), (C,D)    &   7  &     2.8    &    4   &     1.3   &      2    &          4   &              6 \\
     16   &                   (A,B), (A,C), (A,D), (A,E), (B,C), (B,E), (C,D)     &   7  &     2.8    &    4   &     1.3   &      2   &           5   &             10 \\
     17   &                   (A,C), (A,D), (A,E), (B,C), (B,D), (B,E), (C,D)       &  7  &     2.8    &    3   &     1.3   &      2   &           5   &             10 \\
     18   &            (A,B), (A,C), (A,D), (A,E), (B,C), (B,D), (B,E), (C,D)      &  8  &     3.2    &    4   &     1.2   &      2   &           5   &             10 \\
     19   &            (A,B), (A,C), (A,D), (A,E), (B,D), (B,E), (C,D), (C,E)      &   8  &     3.2    &    4   &     1.2   &      2  &            5   &             10 \\
     20   &     (A,B), (A,C), (A,D), (A,E), (B,C), (B,D), (B,E), (C,D), (C,E)     &    9  &     3.6    &    4   &     1.1   &      2  &            5   &             10 \\
     21 & (A,B), (A,C), (A,D), (A,E), (B,C), (B,D), (B,E), (C,D), (C,E), (D,E)  &    10  &     4.0   &     4  &      1.0   &      1    &          5   &             10 \\
\hline
\end{tabular}
}
\caption{Basic properties of the various connected families for graphs with $n=5$ nodes. }
\label{tab:n_5_families_basics}
\end{table*}

\section{Graph Properties and Efficient Evaluation}
We study simple graph properties that can impact the efficiency of QKD networks. 
\subsection{Fundamental Graph Properties}
\label{section_basic_graph_properties}
Consider a connected graph $G=(V,E)$ describing a QKD network. Denote by  $n=|V|$ the number of nodes in a graph. 
\begin{itemize}
    \item Number of edges
    \item Mean degree - The mean degree of a node among the $n$ nodes
    \item Max degree - Maximal degree among the $n$ nodes
    \item Mean distance - The average distance in the graph among all $\binom{n}{2}$ pairs of nodes. Distance affects latency for key distribution. 
    \item Max distance (diameter) - The longest shortest path in the graph represents the worst-case latency for key distribution.
    \item Critical nodes - A critical node refers to a node whose removal (with its connected edges) eliminates the graph connectivity. 
    \item Pairs with Disjoint Paths - Connectivity is reliable when maintained upon failures. We measure the number of pairs of nodes with at least two link-disjoint paths.
    Such paths allow connectivity  even upon any single failure of a link. 
\end{itemize}
Disjoint paths are not only useful for availability under failures.
In trusted-relay QKD networks, they can also support trust reduction via multipath secret sharing; we revisit this idea when defining trust-aware objectives in Sec.~2.3.

\textbf{Loss-weighted edges and QKD-feasible connectivity.}
For analyzing QKD networks, hop count alone is insufficient. Each edge $e\in E$ corresponds to an optical path with total loss (dB)
\begin{equation}
L_e \;=\; \alpha_{\mathrm{fib}}\cdot\ell_e \;+\; IL_e, \nonumber
\end{equation}
where $\alpha_{\mathrm{fib}}\approx 0.2$\,dB/km at 1550\,nm, $\ell_e$ is fiber length, and $IL_e$ aggregates insertion loss (connectors/patch panels, WDM elements, switches). Since SKR can degrade sharply with a few dB of extra loss (Fig.~\ref{fig:attenuation_sensitivity}), we associate to each edge an effective key-generation capacity
\begin{equation}
K_e \;=\; \max\{0, \mathrm{SKR}(L_e)\}. \nonumber
\end{equation}
A simple analytical approximation for the loss dependence can be obtained from the exponential attenuation of photons in optical fibre. 
If $\eta_e = 10^{-L_e/10}$ denotes the channel transmittance, the secret-key rate for common DV QKD protocols (e.g., decoy-state BB84) scales approximately linearly with $\eta_e$ in the moderate-loss regime. 
A convenient first-order model is therefore
\begin{equation}
\mathrm{SKR}(L_e) \;\approx\;
\begin{cases}
K_0 \cdot 10^{-L_e/10}, & L_e \le L_{\max},\\
0, & L_e > L_{\max},
\end{cases}
\nonumber
\end{equation}
where $K_0$ is the achievable SKR at negligible loss (set by source rate, detector efficiency, and post-processing), and $L_{\max}$ is a platform- and protocol-dependent loss budget beyond which dark counts / noise and finite-size effects drive the net SKR to (effectively) zero.

Define the feasible-edge set $E_{\mathrm{feas}}=\{e\in E : K_e>0\}$ and the feasible subgraph $G_{\mathrm{feas}}=(V,E_{\mathrm{feas}})$. We call the network \emph{QKD-feasible connected} if $G_{\mathrm{feas}}$ is connected (or if a target subset of node pairs remains connected in $G_{\mathrm{feas}}$).

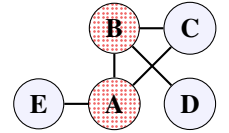
\begin{wrapfigure}{r}{3.8cm}
    \centering
        \begin{tikzpicture}[
    node/.style={
        circle, 
        draw=black, 
        fill=blue!5, 
        minimum size=0.18cm, 
        font=\bfseries
    },
    edge/.style={
        thick, 
        black
    },
    cnode/.style={
        circle, 
        draw=black, 
        pattern={Dots[distance=1.7pt, radius=0.45pt]},
        pattern color=red!60,
        minimum size=0.18cm, 
        font=\bfseries
    },
]

    \node[cnode] (A) at (0,0) {A};

    \node[cnode] (B) at (0,1) {B};
    \node[node] (C) at (1,1)  {C};
    \node[node] (D) at (1,0){D};
    \node[node] (E) at (-1,0){E};

    \draw[edge] (A) -- (B);
    \draw[edge] (A) -- (C);
    \draw[edge] (A) -- (E);
 \draw[edge] (B) -- (C);
    \draw[edge] (B) -- (D);
\end{tikzpicture}
  \caption{Graph example}
    \label{fig:graph_example}
\end{wrapfigure}

\begin{table}[t!]
\centering
\scriptsize
\resizebox{0.5\textwidth}{!}{%
\begin{tabular}{l|cccccc}
\hline
Number of nodes ($n$) & 3 & 4 & 5 & 6 & 7 & 8\\
\hline
Total families & 4 & 11 & 34 & 156 & 1044 & 12346\\
Connected families & 2 & 6 & 21 & 112 & 853 & 11117\\
\hline
\end{tabular}
}
\caption{Number of families of graphs as a function of the number of nodes. }
\label{tab:families_count}
\end{table}

\textbf{A Graph Example.}  
Fig.~\ref{fig:graph_example} illustrates an example of a graph  $G=(V,E)$ with $n=|V|=5$ nodes A, B, C, D, E with $|E|=5$ edges. The mean node degree is directly given by the number of edges as $2 \cdot |E|/n$ = 10/5 = 2 such that the nodes with the maximal degree are A and B, each with degree 3. To compute the mean distance we consider the $\binom{n}{2}=10$ pairs of nodes 
which together imply mean distance of 1.6 and diameter (maximal distance) of 3.

\begin{figure*}[t!]
    \centering
 \subfigure[{\small Star} (1)]{
 \label{fig:family_1}
        \centering

        \begin{tikzpicture}[
        scale=0.85, 
    node/.style={
        circle, 
        draw=black, 
        fill=blue!5, 
        minimum size=0.18cm, 
        font=\bfseries
    },
    edge/.style={
        thick, 
        black
    },
    cnode/.style={
        circle, 
        draw=black, 
        pattern={Dots[distance=1.7pt, radius=0.45pt]},
        pattern color=red!60,
        minimum size=0.18cm, 
        font=\bfseries
    },
]

    \node[cnode] (A) at (0,0) {A};

    \node[node] (B) at (0,1) {B};
    \node[node] (C) at (1,1)  {C};
    \node[node] (D) at (1,0){D};
    \node[node] (E) at (-1,0){E};

    \draw[edge] (A) -- (B);
    \draw[edge] (A) -- (C);
    \draw[edge] (A) -- (D);
    \draw[edge] (A) -- (E);

\end{tikzpicture}
 }
    \hfill
     \subfigure[{\small Fork} (2)]{
    \label{fig:family_2}
        \centering
         \begin{tikzpicture}[
        scale=0.85, 
    node/.style={
        circle, 
        draw=black, 
        fill=blue!5, 
        minimum size=0.18cm, 
        font=\bfseries
    },
    edge/.style={
        thick, 
        black
    },
    cnode/.style={
        circle, 
        draw=black, 
        pattern={Dots[distance=1.7pt, radius=0.45pt]},
        pattern color=red!60,
        minimum size=0.18cm, 
        font=\bfseries
    },
]

    \node[cnode] (A) at (0,0) {A};

    \node[cnode] (B) at (0,1) {B};
    \node[node] (C) at (1,1)  {C};
    \node[node] (D) at (1,0){D};
    \node[node] (E) at (-1,0){E};

    \draw[edge] (A) -- (B);
    \draw[edge] (A) -- (D);
    \draw[edge] (A) -- (E);
    \draw[edge] (B) -- (C);
    
\end{tikzpicture}       
}
    \hfill
     \subfigure[{\small Path} (3)]{
    \label{fig:family_3}
        \centering
         \begin{tikzpicture}[
        scale=0.85, 
    node/.style={
        circle, 
        draw=black, 
        fill=blue!5, 
        minimum size=0.18cm, 
        font=\bfseries
    },
    edge/.style={
        thick, 
        black
    },
    cnode/.style={
        circle, 
        draw=black, 
        pattern={Dots[distance=1.7pt, radius=0.45pt]},
        pattern color=red!60,
        minimum size=0.18cm, 
        font=\bfseries
    },
]

    \node[cnode] (A) at (0,0) {A};

    \node[cnode] (B) at (1,1) {B};
    \node[cnode] (C) at (1,0)  {C};
    \node[node] (D) at (0,1){D};
    \node[node] (E) at (-1,0){E};

    \draw[edge] (A) -- (C);
    \draw[edge] (A) -- (E);
    \draw[edge] (B) -- (C);
    \draw[edge] (B) -- (D);
    
\end{tikzpicture}      
}
    \hfill
     \subfigure[{\small Triangle + Star-center} (4)]{
    \label{fig:family_4}
        \centering
        \begin{tikzpicture}[
        scale=0.85, 
    node/.style={
        circle, 
        draw=black, 
        fill=blue!5, 
        minimum size=0.18cm, 
        font=\bfseries
    },
    edge/.style={
        thick, 
        black
    },
    cnode/.style={
        circle, 
        draw=black, 
        pattern={Dots[distance=1.7pt, radius=0.45pt]},
        pattern color=red!60,
        minimum size=0.18cm, 
        font=\bfseries
    },
]

    \node[cnode] (A) at (0,0) {A};

    \node[node] (B) at (1,1) {B};
    \node[node] (C) at (1,0)  {C};
    \node[node] (D) at (0,1){D};
    \node[node] (E) at (-1,0){E};

    \draw[edge] (A) -- (B);
    \draw[edge] (A) -- (C);
    \draw[edge] (A) -- (D);
    \draw[edge] (A) -- (E);
 \draw[edge] (B) -- (C);
\end{tikzpicture}
 }
    \hfill
     \subfigure[{\small $C_3$ with Path-tail} (5)]{
    \label{fig:family_5}
        \centering
        \begin{tikzpicture}[
        scale=0.85, 
    node/.style={
        circle, 
        draw=black, 
        fill=blue!5, 
        minimum size=0.18cm, 
        font=\bfseries
    },
    edge/.style={
        thick, 
        black
    },
    cnode/.style={
        circle, 
        draw=black, 
        pattern={Dots[distance=1.7pt, radius=0.45pt]},
        pattern color=red!60,
        minimum size=0.18cm, 
        font=\bfseries
    },
]

    \node[cnode] (A) at (0,0) {A};

    \node[cnode] (B) at (0,1) {B};
    \node[node] (C) at (1,1)  {C};
    \node[node] (D) at (1,0){D};
    \node[node] (E) at (-1,0){E};

    \draw[edge] (A) -- (B);
    \draw[edge] (A) -- (C);
    \draw[edge] (A) -- (E);
 \draw[edge] (B) -- (C);
    \draw[edge] (B) -- (D);
\end{tikzpicture}
}
    \hfill
     \subfigure[{$C_4$ with tail} (6)]{
    \label{fig:family_6}
        \begin{tikzpicture}[
                scale=0.85, 
    node/.style={
        circle, 
        draw=black, 
        fill=blue!5, 
        minimum size=0.18cm, 
        font=\bfseries
    },
    edge/.style={
        thick, 
        black
    },
    cnode/.style={
        circle, 
        draw=black, 
        pattern={Dots[distance=1.7pt, radius=0.45pt]},
        pattern color=red!60,
        minimum size=0.18cm, 
        font=\bfseries
    },
]

    \node[cnode] (A) at (0,0) {A};

    \node[node] (B) at (1,1) {B};
    \node[node] (C) at (1,0)  {C};
    \node[node] (D) at (0,1){D};
    \node[node] (E) at (-1,0){E};

    \draw[edge] (A) -- (C);
    \draw[edge] (A) -- (D);
    \draw[edge] (A) -- (E);
 \draw[edge] (B) -- (C);
    \draw[edge] (B) -- (D);
\end{tikzpicture}
}
\\
     \subfigure[{\small $C_3$ with 2 tails (sep)} (7)]{
    \label{fig:family_7}
        \centering
        \begin{tikzpicture}[
                scale=0.85, 
    node/.style={
        circle, 
        draw=black, 
        fill=blue!5, 
        minimum size=0.18cm, 
        font=\bfseries
    },
    edge/.style={
        thick, 
        black
    },
    cnode/.style={
        circle, 
        draw=black, 
        pattern={Dots[distance=1.7pt, radius=0.45pt]},
        pattern color=red!60,
        minimum size=0.18cm, 
        font=\bfseries
    },
]

    \node[node] (A) at (0,0) {A};

    \node[cnode] (B) at (0,1) {B};
    \node[cnode] (C) at (1,1)  {C};
    \node[node] (D) at (1,0){D};
    \node[node] (E) at (-1,0){E};

    \draw[edge] (A) -- (B);
    \draw[edge] (A) -- (E);
    \draw[edge] (B) -- (C);
 \draw[edge] (B) -- (E);
    \draw[edge] (C) -- (D);
\end{tikzpicture}
}
\hfill
     \subfigure[{\small Cycle ($C_5$)} (8)]{
    \label{fig:family_8}
        \centering
        \begin{tikzpicture}[
                scale=0.85, 
    node/.style={
        circle, 
        draw=black, 
        fill=blue!5, 
        minimum size=0.18cm, 
        font=\bfseries
    },
    edge/.style={
        thick, 
        black
    },
    cnode/.style={
        circle, 
        draw=black, 
        pattern={Dots[distance=1.7pt, radius=0.45pt]},
        pattern color=red!60,
        minimum size=0.18cm, 
        font=\bfseries
    },
]

    \node[node] (A) at (0,0) {A};

    \node[node] (B) at (0,1) {B};
    \node[node] (C) at (1,1)  {C};
    \node[node] (D) at (1,0){D};
    \node[node] (E) at (-1,0){E};

    \draw[edge] (A) -- (D);
    \draw[edge] (A) -- (E);
    \draw[edge] (B) -- (C);
 \draw[edge] (B) -- (E);
    \draw[edge] (C) -- (D);
\end{tikzpicture}
}
    \hfill
     \subfigure[{\small Diamond + tail (A)} (9)]{
    \label{fig:family_9}
        \centering
        \begin{tikzpicture}[
                scale=0.85, 
    node/.style={
        circle, 
        draw=black, 
        fill=blue!5, 
        minimum size=0.18cm, 
        font=\bfseries
    },
    edge/.style={
        thick, 
        black
    },
    cnode/.style={
        circle, 
        draw=black, 
        pattern={Dots[distance=1.7pt, radius=0.45pt]},
        pattern color=red!60,
        minimum size=0.18cm, 
        font=\bfseries
    },
]

    \node[cnode] (A) at (0,0) {A};

    \node[node] (B) at (0,1) {B};
    \node[node] (C) at (1,1)  {C};
    \node[node] (D) at (1,0){D};
    \node[node] (E) at (-1,0){E};

    \draw[edge] (A) -- (B);
    \draw[edge] (A) -- (C);
    \draw[edge] (A) -- (D);
 \draw[edge] (A) -- (E);
    \draw[edge] (B) -- (C);
  \draw[edge] (B) -- (D);
\end{tikzpicture}
}
    \hfill
     \subfigure[{\small Bipartite ($K_{2,3}$)} (10)]{
    \label{fig:family_10}
        \centering
        \begin{tikzpicture}[
                scale=0.85, 
    node/.style={
        circle, 
        draw=black, 
        fill=blue!5, 
        minimum size=0.18cm, 
        font=\bfseries
    },
    edge/.style={
        thick, 
        black
    },
    cnode/.style={
        circle, 
        draw=black, 
        pattern={Dots[distance=1.7pt, radius=0.45pt]},
        pattern color=red!60,
        minimum size=0.18cm, 
        font=\bfseries
    },
]

    \node[node] (A) at (0,0) {A};

    \node[node] (B) at (1,0) {B};
    \node[node] (C) at (0,1)  {C};
    \node[node] (D) at (1,1){D};
    \node[node] (E) at (2,1){E};

    \draw[edge] (A) -- (C);
    \draw[edge] (A) -- (D);
    \draw[edge] (A) -- (E);
 \draw[edge] (B) -- (C);
    \draw[edge] (B) -- (D);
    \draw[edge] (B) -- (E);
\end{tikzpicture}
}
\hfill
     \subfigure[{\small Diamond + tail (D)} (11)]{
    \label{fig:family_11}
        \centering
        \begin{tikzpicture}[
                scale=0.85, 
    node/.style={
        circle, 
        draw=black, 
        fill=blue!5, 
        minimum size=0.18cm, 
        font=\bfseries
    },
    edge/.style={
        thick, 
        black
    },
    cnode/.style={
        circle, 
        draw=black, 
        pattern={Dots[distance=1.7pt, radius=0.45pt]},
        pattern color=red!60,
        minimum size=0.18cm, 
        font=\bfseries
    },
]

    \node[cnode] (A) at (0,0) {A};

    \node[node] (B) at (0,1) {B};
    \node[node] (C) at (1,1)  {C};
    \node[node] (D) at (1,0){D};
    \node[node] (E) at (-1,0){E};
    \draw[edge] (A) -- (B);
    \draw[edge] (A) -- (C);
    \draw[edge] (A) -- (E);
 \draw[edge] (B) -- (C);
    \draw[edge] (B) -- (D);
    \draw[edge] (C) -- (D);
\end{tikzpicture}
}
    \hfill
     \subfigure[{Butterfly} (12)]{
    \label{fig:family_12}
        \centering
        \begin{tikzpicture}[
                scale=0.85, 
    node/.style={
        circle, 
        draw=black, 
        fill=blue!5, 
        minimum size=0.18cm, 
        font=\bfseries
    },
    edge/.style={
        thick, 
        black
    },
    cnode/.style={
        circle, 
        draw=black, 
        pattern={Dots[distance=1.7pt, radius=0.45pt]},
        pattern color=red!60,
        minimum size=0.18cm, 
        font=\bfseries
    },
]

    \node[cnode] (A) at (0,0) {A};

    \node[node] (B) at (0,1) {B};
    \node[node] (C) at (1,1)  {C};
    \node[node] (D) at (1,0){D};
    \node[node] (E) at (-1,0){E};

    \draw[edge] (A) -- (B);
    \draw[edge] (A) -- (C);
    \draw[edge] (A) -- (D);
      \draw[edge] (A) -- (E);  
 \draw[edge] (B) -- (E);
    \draw[edge] (C) -- (D);
\end{tikzpicture}
}
\hfill
     \subfigure[{\small $C_5$ + chord}  (13)]{
    \label{fig:family_13}
        \centering
        \begin{tikzpicture}[
                scale=0.85, 
    node/.style={
        circle, 
        draw=black, 
        fill=blue!5, 
        minimum size=0.18cm, 
        font=\bfseries
    },
    edge/.style={
        thick, 
        black
    },
    cnode/.style={
        circle, 
        draw=black, 
        pattern={Dots[distance=1.7pt, radius=0.45pt]},
        pattern color=red!60,
        minimum size=0.18cm, 
        font=\bfseries
    },
]

    \node[node] (A) at (0,0) {A};

    \node[node] (B) at (0,1) {B};
    \node[node] (C) at (1,1)  {C};
    \node[node] (D) at (1,0){D};
    \node[node] (E) at (-1,0){E};

     \draw[edge] (A) -- (B);
    \draw[edge] (A) -- (D);
    \draw[edge] (A) -- (E);
    \draw[edge] (B) -- (C);
 \draw[edge] (B) -- (E);
    \draw[edge] (C) -- (D);
\end{tikzpicture}
  }
    \hfill
     \subfigure[	$K_4$ with tail (center)  (14)]{
    \label{fig:family_14}
        \centering
        \begin{tikzpicture}[
                scale=0.85, 
    node/.style={
        circle, 
        draw=black, 
        fill=blue!5, 
        minimum size=0.18cm, 
        font=\bfseries
    },
    edge/.style={
        thick, 
        black
    },
    cnode/.style={
        circle, 
        draw=black, 
        pattern={Dots[distance=1.7pt, radius=0.45pt]},
        pattern color=red!60,
        minimum size=0.18cm, 
        font=\bfseries
    },
]

    \node[node] (A) at (0,0) {A};

    \node[node] (B) at (0,1) {B};
    \node[node] (C) at (1,1)  {C};
    \node[node] (D) at (1,0){D};
    \node[node] (E) at (-1,0){E};

    \draw[edge] (A) -- (B);
    \draw[edge] (A) -- (C);
    \draw[edge] (A) -- (D);
 \draw[edge] (A) -- (E);
    \draw[edge] (B) -- (C);
  \draw[edge] (B) -- (D);
  \draw[edge] (B) -- (E);
\end{tikzpicture}
}
\hfill
     \subfigure[	{\small $K_4$ with tail (leaf)}  (15)]{
    \label{fig:family_15}
        \begin{tikzpicture}[
                scale=0.85, 
    node/.style={
        circle, 
        draw=black, 
        fill=blue!5, 
        minimum size=0.18cm, 
        font=\bfseries
    },
    edge/.style={
        thick, 
        black
    },
    cnode/.style={
        circle, 
        draw=black, 
        pattern={Dots[distance=1.7pt, radius=0.45pt]},
        pattern color=red!60,
        minimum size=0.18cm, 
        font=\bfseries
    },
]

    \node[cnode] (A) at (0,0) {A};

    \node[node] (B) at (0,1) {B};
    \node[node] (C) at (1,1)  {C};
    \node[node] (D) at (1,0){D};
    \node[node] (E) at (-1,0){E};
     \draw[edge] (A) -- (B);
    \draw[edge] (A) -- (C);
    \draw[edge] (A) -- (D);
    \draw[edge] (A) -- (E);
 \draw[edge] (B) -- (C);
    \draw[edge] (B) -- (D);
    \draw[edge] (C) -- (D);
\end{tikzpicture}
}
\hfill
     \subfigure[	{\small $C_5$ with 2 chords (X)}  (16)]{
    \label{fig:family_16}
        \begin{tikzpicture}[
                scale=0.85, 
    node/.style={
        circle, 
        draw=black, 
        fill=blue!5, 
        minimum size=0.18cm, 
        font=\bfseries
    },
    edge/.style={
        thick, 
        black
    },
    cnode/.style={
        circle, 
        draw=black, 
        pattern={Dots[distance=1.7pt, radius=0.45pt]},
        pattern color=red!60,
        minimum size=0.18cm, 
        font=\bfseries
    },
]

    \node[node] (A) at (0,0) {A};

    \node[node] (B) at (0,1) {B};
    \node[node] (C) at (1,1)  {C};
    \node[node] (D) at (1,0){D};
    \node[node] (E) at (-1,0){E};

    \draw[edge] (A) -- (B);
    \draw[edge] (A) -- (C);
    \draw[edge] (A) -- (D);
    \draw[edge] (A) -- (E);
 \draw[edge] (B) -- (C);
    \draw[edge] (B) -- (E);
    \draw[edge] (C) -- (D);
\end{tikzpicture}
 }
    \hfill
     \subfigure[	{\small $C_5$ with 2 chords (parallel)}  (17)]{
    \label{fig:family_17}
        \begin{tikzpicture}[
                scale=0.85, 
    node/.style={
        circle, 
        draw=black, 
        fill=blue!5, 
        minimum size=0.18cm, 
        font=\bfseries
    },
    edge/.style={
        thick, 
        black
    },
    cnode/.style={
        circle, 
        draw=black, 
        pattern={Dots[distance=1.7pt, radius=0.45pt]},
        pattern color=red!60,
        minimum size=0.18cm, 
        font=\bfseries
    },
]

    \node[node] (A) at (0,0) {A};

    \node[node] (B) at (0,1) {B};
    \node[node] (C) at (1,0)  {C};
    \node[node] (D) at (1,1){D};
    \node[node] (E) at (-1,0){E};


    \draw[edge] (A) -- (C);
    \draw[edge] (A) -- (D);
      \draw[edge] (A) -- (E);  
 \draw[edge] (B) -- (D);
     \draw[edge] (B) -- (E);
    \draw[edge] (C) -- (D);
\end{tikzpicture}
}
\hfill
     \subfigure[	{\small $K_5$ minus 2 adjacent edges}  (18)]{
    \label{fig:family_18}
        \centering
        \begin{tikzpicture}[
                scale=0.85, 
    node/.style={
        circle, 
        draw=black, 
        fill=blue!5, 
        minimum size=0.18cm, 
        font=\bfseries
    },
    edge/.style={
        thick, 
        black
    },
    cnode/.style={
        circle, 
        draw=black, 
        pattern={Dots[distance=1.7pt, radius=0.45pt]},
        pattern color=red!60,
        minimum size=0.18cm, 
        font=\bfseries
    },
]

    \node[node] (A) at (0,0) {A};

    \node[node] (B) at (0,1) {B};
    \node[node] (C) at (1,1)  {C};
    \node[node] (D) at (1,0){D};
    \node[node] (E) at (-1,0){E};

     \draw[edge] (A) -- (B);
     \draw[edge] (A) -- (C);
    \draw[edge] (A) -- (D);
    \draw[edge] (A) -- (E);
    \draw[edge] (B) -- (C);
    \draw[edge] (B) -- (D);
 \draw[edge] (B) -- (E);
    \draw[edge] (C) -- (D);
\end{tikzpicture}
}
    \hfill
     \subfigure[	Wheel ($W_5$)  (19)]{
    \label{fig:family_19}
        \centering
        \begin{tikzpicture}[
                scale=0.85, 
    node/.style={
        circle, 
        draw=black, 
        fill=blue!5, 
        minimum size=0.18cm, 
        font=\bfseries
    },
    edge/.style={
        thick, 
        black
    },
    cnode/.style={
        circle, 
        draw=black, 
        pattern={Dots[distance=1.7pt, radius=0.45pt]},
        pattern color=red!60,
        minimum size=0.18cm, 
        font=\bfseries
    },
]

    \node[node] (A) at (0,0) {A};

    \node[node] (B) at (0,1) {B};
    \node[node] (C) at (1,1)  {C};
    \node[node] (D) at (1,0){D};
    \node[node] (E) at (-1,0){E};

    \draw[edge] (A) -- (B);
    \draw[edge] (A) -- (C);
    \draw[edge] (A) -- (D);
 \draw[edge] (A) -- (E);

  \draw[edge] (B) -- (D);
  \draw[edge] (B) -- (E);
    \draw[edge] (C) -- (D);
        \draw[edge] (C) -- (E);
\end{tikzpicture}
}
\hfill
     \subfigure[	$K_5$ minus 1 edge  (20)]{
    \label{fig:family_20}
        \centering
        \begin{tikzpicture}[
                scale=0.85, 
    node/.style={
        circle, 
        draw=black, 
        fill=blue!5, 
        minimum size=0.18cm, 
        font=\bfseries
    },
    edge/.style={
        thick, 
        black
    },
    cnode/.style={
        circle, 
        draw=black, 
        pattern={Dots[distance=1.7pt, radius=0.45pt]},
        pattern color=red!60,
        minimum size=0.18cm, 
        font=\bfseries
    },
]

    \node[node] (A) at (0,0) {A};

    \node[node] (B) at (0,1) {B};
    \node[node] (C) at (1,1)  {C};
    \node[node] (D) at (1,0){D};
    \node[node] (E) at (-1,0){E};
   
     \draw[edge] (A) -- (B);
    \draw[edge] (A) -- (C);
    \draw[edge] (A) -- (D);
    \draw[edge] (A) -- (E);
 \draw[edge] (B) -- (C);
    \draw[edge] (B) -- (D);
    \draw[edge] (B) -- (E);
    \draw[edge] (C) -- (D);
    \draw[edge] (C) -- (E);
\end{tikzpicture}
}
\hfill
     \subfigure[	{\small Complete Graph ($K_5$)}  (21)]{
    \label{fig:family_21}
        \centering
        \begin{tikzpicture}[
                scale=0.85, 
    node/.style={
        circle, 
        draw=black, 
        fill=blue!5, 
        minimum size=0.18cm, 
        font=\bfseries
    },
    edge/.style={
        thick, 
        black
    }
]

    \node[node] (A) at (0,0) {A};

    \node[node] (B) at (0,1) {B};
    \node[node] (C) at (1,1)  {C};
    \node[node] (D) at (1,0){D};
    \node[node] (E) at (-1,0){E};

     \draw[edge] (A) -- (B);
    \draw[edge] (A) -- (C);
    \draw[edge] (A) -- (D);
    \draw[edge] (A) -- (E);
 \draw[edge] (B) -- (C);
    \draw[edge] (B) -- (D);
    \draw[edge] (B) -- (E);
    \draw[edge] (C) -- (D);
    \draw[edge] (C) -- (E);
 \draw[edge] (D) to [out=210, in=330] (E);
    
\end{tikzpicture}
  }
    \caption{Illustration of the families of graphs with $n=5$ nodes. The properties of the families are detailed in Table~\ref{tab:n_5_families_basics}. Critical nodes are shown in red with dots.}
    \label{fig:visual_illustration_of_families}
\end{figure*}
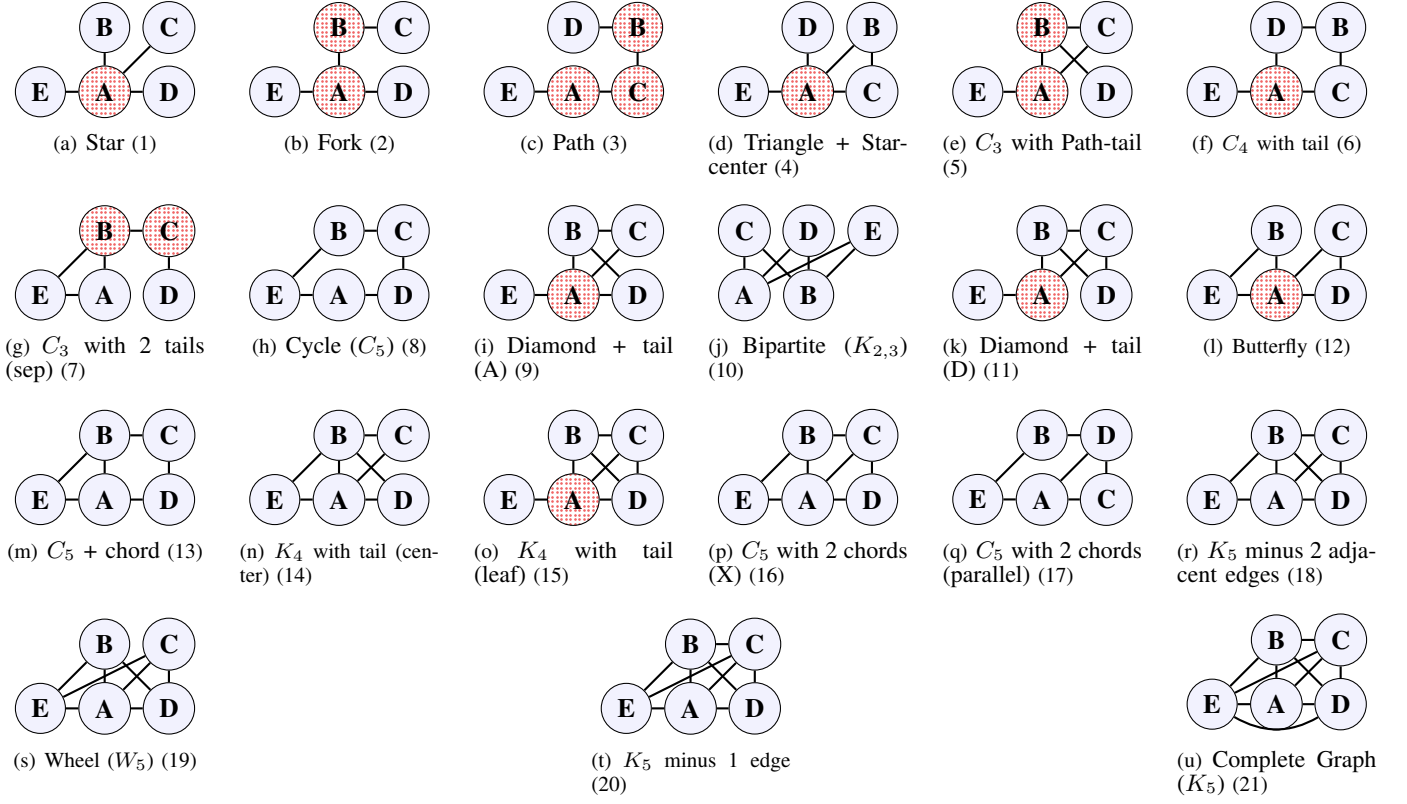

We can see that connectivity is maintained even if one of the nodes C, D or E are eliminated but not if A or B are disconnected. We refer to A and B as critical nodes and illustrate them as being colored in red with dots. We can also see that among the pair of nodes, between any two nodes among A, B, C we can establish two link-disjoint paths. In this case, they refer to the directed edge and to the path through the other nodes among the three. We cannot establish such paths for pairs that involve D and E so the number of pairs with disjoint paths is 3.

\subsection{Practical Evaluation for Small Graphs}
For a small number of nodes (such as $n \le 8$), we would like to simply identify graphs with properties allowing efficient QKD. However, examining all potential graphs can be intensive. As there are $m$ = $n \cdot$ ($n-1$) / 2 edges, and each edge can be included or not, there are $2^m$ potential graphs. 
As the mentioned properties from Sec.~\ref{section_basic_graph_properties} can be derived from the graph structure,  many of those graphs are identical with respect to these properties. 
There is no need to consider all such graphs but only to examine a representative of each group of graphs with the same structure.

We \emph{refer to a group of graphs with the same structure as a family}, namely we see nodes as without any particular labeling. 
These families are also known as isomorphism classes of unlabeled graphs. 
For instance, with $n=3$ and nodes A, B, C we refer to the three graphs with two edges {(A,B), (A,C)} or {(A,B), (B,C)} or {(A,C), (B,C)} 
as of the same family.

In the same family, graphs have the same basic properties such as number of edges, average and max degree, etc. Interestingly, for small $n$, the number of families is not that large. Among these families, we can focus on connected families, namely those referring to connected graphs. For $n=5$ for instance, there are only 21 connected families. Table~\ref{tab:families_count} shows the number of families and the number of connected families per various values of the nodes count $n \in \{3, 4, 5, 6, 7, 8\}$.

Consider, for instance, graphs with $n=5$ nodes. We can compare the 21 connected families of graphs with regard to several properties affecting QKD network performance. 

Table~\ref{tab:n_5_families_basics} shows several graph properties that can be computed for the various 21 families of graphs which are connected and have $n=5$ nodes. It refers to the  properties from Sec.~\ref{section_basic_graph_properties}. 
Families are arranged in an increasing order of the number of edges (ranging between 4 to 10). 
The graph from Fig.~\ref{fig:graph_example} appears in the table as the family with index (5). 

There is a clear connection between several of the properties. As many of these properties have been well studied in graph theory, we present only a short overview. 

(i) Upon adding edges, a node can become non-critical.  

(ii) A node with degree 1 is a non-critical node. However, such a node cannot be among the pairs of nodes with two or more disjoint paths. 

(iii) There are graphs with $n$ edges in the form of a cycle (often denoted by $C_n$) where all nodes are non-critical and all pairs of nodes have disjoint paths. There are no graphs with less than $n$ edges having any of these two properties. 

Fig.~\ref{fig:visual_illustration_of_families} provides a visual illustration of these 21 families, previously summarized in Table~\ref{tab:n_5_families_basics}. Again, the critical nodes  are shown in red with dots. We informally refer to the structure of each family in the caption of each subfigure. 
\subsection{Cost Functions for Graphs}\label{section_cost_functions}
There is no one way to measure the various properties of a graph $G$. Accordingly, we do not aim to indicate a particular formula and only overview two such forms that can be justified. Before presenting the formulas, we overview their potential components: 

\emph{O} - This is the operational cost of the graphs, measured as its number of nodes plus its number of edges.
\begin{equation}
    O = \alpha \cdot |V| + |E| = \alpha \cdot n + m \nonumber 
\end{equation}
The parameter $\alpha$ allows to change the weight of overhead of nodes with regards to links.  We typically set $\alpha = 2$. When the number of nodes $n$ is fixed, the operational cost is affected only by the number of edges.  

\emph{R} - The reliability factor of the graph can take into account both properties of the number of non-critical nodes and the number of pairs with link-disjoint paths. Let $n_{nc} \in [0,n]$ be the number of non-critical nodes. Likewise, let $n_p \in [0, \binom{n}{2}]$ the number of pairs of nodes with link-disjoint pairs of paths. For some weight $\beta$, the factor is given as  \begin{equation}
    R = \frac{1}{\beta + 1} \Big(\beta \cdot n_{nc} / n + n_p /   \binom{n}{2} \Big) . \nonumber 
\end{equation}
The reliability factor has values in $[0,1]$. The parameter $\beta$ allows to change the weight of the two components of the factor. We typically set $\beta = 2$ such that the weight of the ratio of non-critical nodes has a higher weight than the ratio of pairs with disjoint paths. 

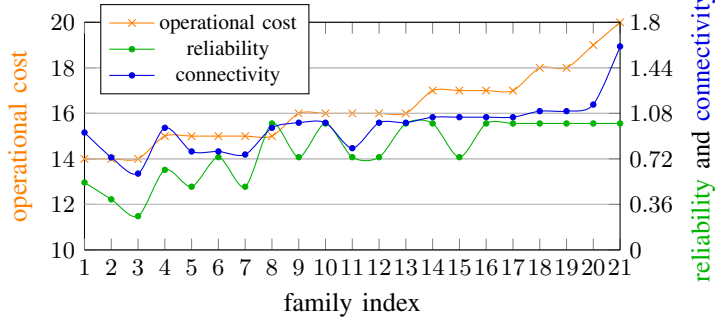
\begin{figure}[t!]
\centering
\begin{tikzpicture}
\pgfplotsset{
    scale only axis,
    xmin=1, xmax=21, 
    xtick = {1, 2, 3, 4, 5, 6, 7, 8, 9, 10, 11, 12, 13, 14, 15, 16, 17, 18, 19, 20,21},
        tick label style={font=\small},
   legend style={nodes={scale=0.8}, at={(0.42,1.08)}}, 
    width = 0.80 \columnwidth,
    height = 0.34 \columnwidth,
    ymajorgrids,
}
\begin{axis}[
  axis y line*=left,
  axis x line=none,
     ymin=10, ymax=20,
   ytick = {10, 12, 14, 16, 18, 20}, 
  ylabel=\textcolor{orange}{operational cost}
]
\addplot[smooth,mark=x,orange]
  coordinates{
  (1,14)
  (2,14)
  (3,14)
  (4,15)
  (5,15)
  (6,15)
  (7,15)
  (8,15)
  (9,16)
  (10,16)
  (11,16)
  (12,16)
  (13,16)
  (14,17)
  (15,17)
  (16,17)
  (17,17)
  (18,18)
  (19,18)
  (20,19)
  (21,20)
}; \label{plot_one}
\end{axis}
\begin{axis}[
  axis y line*=right,
ymin=0, ymax=1.8,
   ytick = {0, 0.36, 0.72, 1.08, 1.44, 1.8}, 
  xlabel= family index,
  ylabel={\textcolor{green}{reliability} and \textcolor{blue}{connectivity}},
]
\addlegendimage{/pgfplots/refstyle=plot_one}\addlegendentry{operational cost}
\addplot[smooth,mark=*,mark size=1pt, green]
  coordinates{ 
(1, 0.5333333333333333)
(2, 0.39999999999999997)
(3, 0.26666666666666666)
(4, 0.6333333333333334)
(5, 0.5)
(6, 0.7333333333333334)
(7, 0.5)
(8, 1.0)
(9, 0.7333333333333334)
(10, 1.0)
(11, 0.7333333333333334)
(12, 0.7333333333333334)
(13, 1.0)
(14, 1.0)
(15, 0.7333333333333334)
(16, 1.0)
(17, 1.0)
(18, 1.0)
(19, 1.0)
(20, 1.0)
(21, 1.0)
}; \addlegendentry{reliability}
\addplot[smooth,mark=*,mark size=1pt, blue]
  coordinates{ 
(1, 0.9285218725581347)
(2, 0.7315626874700456)
(3, 0.6035392171627876)
(4, 0.9656627474604601)
(5, 0.7787602802100485)
(6, 0.7787602802100485)
(7, 0.7544240214534844)
(8, 0.9656627474604601)
(9, 1.0058986952713127)
(10, 1.0058986952713127)
(11, 0.8047189562170501)
(12, 1.0058986952713127)
(13, 1.0058986952713127)
(14, 1.049633421152674)
(15, 1.049633421152674)
(16, 1.049633421152674)
(17, 1.049633421152674)
(18, 1.0973440312050682)
(19, 1.0973440312050682)
(20, 1.1495985088815002)
(21, 1.6094379124341003)
}; \addlegendentry{connectivity}
\end{axis}

\end{tikzpicture}
\caption{Components of the cost function for the 21 graph families connecting $n=5$ nodes (with parameter values $\alpha = \beta = \gamma = 2$). }
\label{figure_cost_distribution_per_family}
\end{figure}

\emph{C} - The connectivity factor takes into account both mean distance among nodes (denoted as $\bar{d}$) and the graph diameter (denoted as $D$). Both are measured in units of number of hops. To  compare graphs of various number of nodes, we scale the values by $\log n$ as often several properties related to distance and diameter in random graphs scale with such a factor under some assumptions~\cite{albert2002statistical, watts1998collective}. 
\begin{equation}
C = \big((\gamma+1) \cdot \log n \big)/ \big(\gamma \cdot \bar{d} + D \big). \nonumber 
\end{equation}

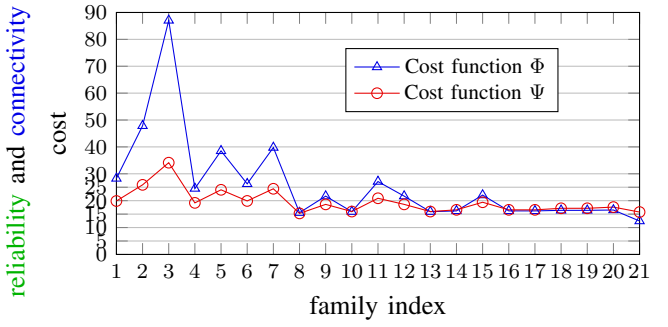
\begin{figure}[t!]
    \centering
\begin{tikzpicture}
\begin{axis}
[ 
    xlabel={family index},
    ylabel = {cost},
    xmin=1, xmax=21,
    ymin = 0, ymax=90,
    xtick = {1, 2, 3, 4, 5, 6, 7,8, 9,10, 11, 12, 13, 14, 15, 16, 17, 18, 19, 20, 21}, 
    ymajorgrids,
    ytick={0, 5, 10, 15, 20, 25, 30, 40, 50, 60, 70, 80, 90},
    width = 0.96 \columnwidth,
    height = 0.54 \columnwidth,
    legend style={nodes={scale=0.80}, at={(0.845,0.853)}}, 
    tick label style={font=\footnotesize} 
] 
\addplot[color=blue, mark=triangle]
coordinates{
(1, 28.2713904975)
(2, 47.8427814643)
(3, 87.0515104445)
(4, 24.5269785084)
(5, 38.5227656209)
(6, 26.2655220142)
(7, 39.7654350106)
(8, 15.5333887220)
(9, 21.6881260759)
(10, 15.9061743890)
(11, 27.1101575949)
(12, 21.6881260759)
(13, 15.9061743890)
(14, 16.1961306346)
(15, 22.0856326836)
(16, 16.1961306346)
(17, 16.1961306346)
(18, 16.4032422502)
(19, 16.4032422502)
(20, 16.5275092348)
(21, 12.4267000000)
};
\addlegendentry{Cost function $\Phi$};
\addplot[color=red, mark=o]
coordinates{
(1, 19.8732360527)
(2, 25.8821943801)
(3, 34.1275306871)
(4, 19.1808285511)
(5, 24.0383324001)
(6, 19.8988636402)
(7, 24.4231649313)
(8, 15.2631526615)
(9, 18.6215758410)
(10, 15.9530401490)
(11, 20.8938210103)
(12, 18.6215758410)
(13, 15.9530401490)
(14, 16.5910408453)
(15, 19.4140084474)
(16, 16.5910408453)
(17, 16.5910408453)
(18, 17.1657366373)
(19, 17.1657366373)
(20, 17.6111162479)
(21, 15.7651036000)
};
\addlegendentry{Cost function $\Psi$};
\end{axis}
\end{tikzpicture}
\caption{Values the cost functions for the 21 graph families connecting $n=5$ nodes (with parameter values $\alpha = \beta = \gamma = 2$). }
\label{figure_cost_values_per_family}
\end{figure}

 The parameter $\gamma$ balances the two values. We typically set $\gamma=2$ such that the mean distance has higher weight than the diameter as it is  influenced by all pairs of nodes. 
 
The three factors $O, R, C$ can be jointly considered in multiple ways. Overall, we prefer lower operational cost and higher reliability and connectivity factors. Two simple cost functions are  
\begin{equation}
\Phi(G) = O/ (R \cdot C) \quad \text{and} \quad \Psi(G) = O / \sqrt{R \cdot C}.  \nonumber 
\end{equation} 

{\bf From Graph Structure to Delivered Secret-Key Rate.}
To connect $\bar d$ and $|E|$ to key delivery, consider an idealized setting where each edge can generate $K$ secret bits/sec and end-to-end key transport uses hop-by-hop one-time-pad protection over trusted relays. 
If all $\binom{n}{2}$ pairs demand the same end-to-end key rate $k$ and traffic is perfectly balanced, the network must ``spend'' approximately $\binom{n}{2}\cdot k\cdot \bar d$ link-key bits/sec (each delivered bit traverses $\bar d$ hops on average). 
Since the network produces at most $|E| \cdot K$ bits/sec across all links, feasibility requires
\begin{equation}
k \le |E|\cdot K\cdot \bar d^{-1} /\binom{n}{2} .\nonumber 
\end{equation}
This highlights why (i) more edges increase aggregate key-generation resources, while (ii) smaller mean distance improves key-transport efficiency. 
In a real QKD network, $K$ becomes edge-dependent ($K\to K_e^{\mathrm{eff}}$) and decays with loss (Fig.~\ref{fig:attenuation_sensitivity}), motivating loss-aware evaluation beyond $\Phi,\Psi$.

Fig.~\ref{figure_cost_distribution_per_family} illustrates the values of the three components (operational, reliability and connectivity) for the 21 families of graphs with $n=5$ nodes. Similarly, Fig.~\ref{figure_cost_values_per_family} shows the values of the two cost functions $\Phi(G)$ and $\Psi(G)$ for those 21 families.  

\begin{table}[t!]
\centering
\scriptsize
\resizebox{0.95\columnwidth}{!}{%
\begin{tabular}{c|cc|}
\hline
Rank & Function $\Phi$ & Function $\Psi$ \\
\hline
21st (Lowest) & Family 21 (12.42) & Family 8 (15.26)  \\
20th & Family 8 (15.53)	& Family 21 (15.76)  \\
19th	& Family 10 (15.90) &	Family 10 (15.95)  \\
\hline
\hline
1st (Highest) & Family 3 (87.05) & Family 3 (34.12)  \\
2nd & Family 2 (47.84)	& Family 2 (25.88)  \\
3rd	& Family 7 (39.76) &	Family 7 (24.42)  \\
\hline
\end{tabular}
}
\caption{Ranking of graph families by their cost values. For the two functions, the three families with the  lowest and highest cost values are shown. Low function values represent efficiency and indicate low operational cost with high reliability and connectivity. }
\label{tab:families_cost_ranking}
\end{table}

Table~\ref{tab:families_cost_ranking} lists the three families with the  lowest  and highest values for the two cost functions. Low values  indicate high efficiency. 
For both cost functions, Family 3 which refers to the path graph has the highest cost. The reason is its low reliability and connectivity values. As shown by the nodes in red in Fig.~\ref{fig:visual_illustration_of_families}, it has more critical nodes than all other families and it also has no pairs with disjoint paths. The path structure also implies the highest mean distance as well as the highest diameter. Family 2 and Family 7 also have relatively high values of the two cost functions. 
On the positive side, the three families with the lowest values for the two cost functions are Family 8, 10, and 21, while the superiority alternates between Family 8 and 21.

\paragraph{QKD-aware extensions (capacity, loss, and trust).}
The structural objectives $\Phi(G), \Psi(G)$ are useful, but a deployed QKD network is naturally a \emph{weighted} graph. Using the per-edge loss $L_e$ and key-generation capacity $K_e$ defined in Sec.~\ref{section_basic_graph_properties}, we obtain two complementary objective families:

\emph{(1) Capacity-aware objectives.} Given a demand matrix $\{d_{st}\}$ (target key supply between node pairs) and per-edge capacities $\{K_e\}$, one can optimize a multi-commodity flow of key material subject to $\sum_{(s,t)} f_{st}(e) \le K_e$ for all edges. A simple objective is to maximize a utility such as $\sum_{s<t} w_{st}\cdot\log(k_{st}+\varepsilon)$, where $k_{st}$ is delivered end-to-end key rate and $w_{st}\ge 0$ is an application-defined weight (priority) for the pair $(s,t)$.

\emph{(2) Trust/risk-aware objectives.} If intermediate nodes are treated as trusted relays, then each multi-hop path induces an explicit trust surface. One can assign each node a risk score $r_v$ (from operational security assessment) and penalize paths by $\mathrm{Risk}(P)=\sum_{v\in P\setminus\{s,t\}} r_v$, or prefer solutions that keep all high-value flows within a low-risk subgraph. Combining (1) and (2) yields a more realistic ``cost function'' for topology design than purely structural metrics.

\subsection{Efficient Graphs for Larger $n$ Values}
For larger values of $n$ (e.g., $n > 8$), exhaustive enumeration of graph families becomes more challenging. A natural next step could be to apply local search and dedicated heuristics which directly optimize the selected cost function (structural $\Phi$, capacity-aware, or trust-aware) under constraints on maximum degree and total number of links. 

We describe two alternative simple methods to construct efficient graphs for such $n$ values:

\textbf{Construction I.}  If for instance $n = k \cdot n_0$ for small $k \in [3,n_0]$, we consider $k$ copies of an efficient family for $n_0$ nodes. We select a representative node from each family as a node with link-disjoint path to many other nodes. We make all representatives to be fully connected by adding $\binom{k}{2}$ links.

\begin{figure}[t!]
    \centering
        \begin{tikzpicture}[
                scale=0.76, 
    node/.style={
        circle, 
        draw=black, 
        fill=blue!5, 
        minimum size=0.14cm, 
        font=\bfseries
    },
    edge/.style={
        thick, 
        black
    },
    cnode/.style={
        circle, 
        draw=black, 
        pattern={Dots[distance=1.7pt, radius=0.45pt]},
        pattern color=red!60,
        minimum size=0.14cm, 
        font=\bfseries
    },
]

    \node[node] (A) at (3,0) {A};

    \node[node] (B) at (4,0) {B};
    \node[node] (C) at (3,1)  {C};
    \node[node] (D) at (4,1){D};
    \node[node] (E) at (5,1){E};

    \node[node] (F) at (-2,0) {F};
    \node[node] (G) at (-3,0) {G};
    \node[node] (H) at (-2,1)  {H};
    \node[node] (I) at (-3,1){I};
    \node[node] (J) at (-4,1){J};
    
    \draw[edge] (A) -- (C);
    \draw[edge] (A) -- (D);
    \draw[edge] (A) -- (E);
 \draw[edge] (B) -- (C);
    \draw[edge] (B) -- (D);
    \draw[edge] (B) -- (E);

    \draw[edge] (F) -- (H);
    \draw[edge] (F) -- (I);
    \draw[edge] (F) -- (J);
 \draw[edge] (G) -- (H);
    \draw[edge] (G) -- (I);
    \draw[edge] (G) -- (J);

    \node[node] (M) at (1.5,0) {M};
    \node[node] (N) at (0.5,0) {N};
    \node[node] (O) at (1.5,1)  {O};
    \node[node] (P) at (0.5,1){P};
    \node[node] (Q) at (-0.5,1){Q};

    \draw[edge] (M) -- (O);
    \draw[edge] (M) -- (P);
    \draw[edge] (M) -- (Q);
 \draw[edge] (N) -- (O);
    \draw[edge] (N) -- (P);
    \draw[edge] (N) -- (Q);

\draw[edge] (A) to[out=200, in=330] (F);
   \draw[edge] (A) -- (M);
      \draw[edge] (M) to[out=220, in=345] (F);
\draw[edge, color=red] (C) to[out=140, in=30] (H);
   \draw[edge, color=red] (C) -- (O);
      \draw[edge,color=red] (O) to[out=140, in=25] (H);
\end{tikzpicture}
    \caption{An illustration of the construction of a graph with $n=15$ nodes based on $k=3$ graphs of an efficient family  with 5 nodes and 6 edges (Family 10).}
    \label{fig:graph_merge_example}
\end{figure}
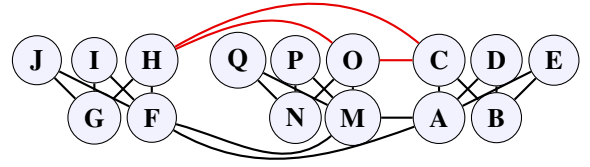

Fig.~\ref{fig:graph_merge_example} illustrates a construction of a graph with $n=15$ nodes based on three graphs of 5 nodes, each in the form of Family 10 in Fig.~\ref{fig:visual_illustration_of_families}, which was ranked  among the families with lowest values of the cost functions (Table~\ref{tab:families_cost_ranking}). In this family, all five nodes are non-critical and there are link-disjoint paths between all 10 pairs. Such properties allow establishing link-disjoint paths between any pair among the $\binom{15}{2}=105$ pairs in the joint topology. All nodes besides the three representing nodes A, F, M are non-critical nodes. Adding the three edges in red makes all nodes be non-critical. 

We deduce interesting properties for Construction I (with the red edges):
\begin{property} Consider Construction I.  
(i) While the links connecting the multiple copies of the family increase the mean node degree by less than 1, they bound the diameter to be at most $2D+1$, where $D=2$ is the diameter of Family 10. 

(ii) The construction allows establishing pairs of link-disjoint paths also between all pairs of nodes from two distinct copies of the $n$-node graph. 

(iii) Moreover, in the construction all nodes are non-critical. 
\end{property}

\begin{proof} 
To see (i), the inequality  $\binom{k}{2} < k^2 / 2 \le k \cdot n_0 / 2 = n / 2$ bounds the increase in the mean node degree as we connect all pairs of copies of $n_0$ nodes with two links, making the added links to be  $2 \cdot \binom{k}{2} < n$.  In each copy of the family nodes can connect within a distance of the diameter $D$. To connect nodes in various copies, the path can use a single link connecting the copies in addition of the use of at most $D$ edges within each family.
For (ii) link-disjoint paths can be established based on the ability to establish such path in Family 10. In each copy we connect the node to the two representing nodes by link-disjoint paths. Family 10 allows establishing such paths. We add to the paths the two links connecting the two pairs of nodes with one node in each copy in each pair. 
To see (iii) recall that all nodes in Family 10 are non-critical. Connecting each pair of copies with two edges guaranties the connectivity is maintained for any failure of a node.  
\end{proof}

\textbf{Construction II.} This construction is hierarchical and is based on an efficient family of $n_0$ nodes for establishing graphs of $(n_0)^2$ nodes. To do so, we replace each node in the family by a copy of the family. Namely, we connect a representative node from $n_0$ copies of the family in a graph with the form of the family itself.

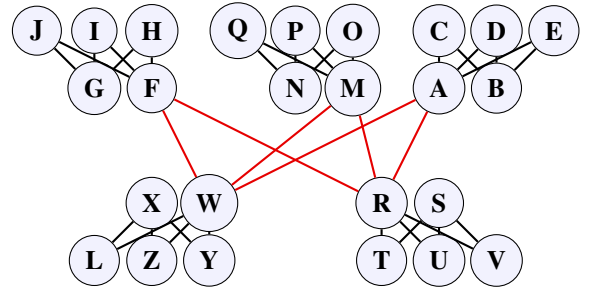
\begin{figure}[t!]
    \centering
        \begin{tikzpicture}[
                scale=0.76, 
    node/.style={
        circle, 
        draw=black, 
        fill=blue!5, 
        minimum size=0.14cm, 
        font=\bfseries
    },
    edge/.style={
        thick, 
        black
    },
    cnode/.style={
        circle, 
        draw=black, 
        pattern={Dots[distance=1.7pt, radius=0.45pt]},
        pattern color=red!60,
        minimum size=0.14cm, 
        font=\bfseries
    },
]

    \node[node] (A) at (3,0) {A};

    \node[node] (B) at (4,0) {B};
    \node[node] (C) at (3,1)  {C};
    \node[node] (D) at (4,1){D};
    \node[node] (E) at (5,1){E};

    \node[node] (F) at (-2,0) {F};
    \node[node] (G) at (-3,0) {G};
    \node[node] (H) at (-2,1)  {H};
    \node[node] (I) at (-3,1){I};
    \node[node] (J) at (-4,1){J};
    
    \draw[edge] (A) -- (C);
    \draw[edge] (A) -- (D);
    \draw[edge] (A) -- (E);
 \draw[edge] (B) -- (C);
    \draw[edge] (B) -- (D);
    \draw[edge] (B) -- (E);

    \draw[edge] (F) -- (H);
    \draw[edge] (F) -- (I);
    \draw[edge] (F) -- (J);
 \draw[edge] (G) -- (H);
    \draw[edge] (G) -- (I);
    \draw[edge] (G) -- (J);

    \node[node] (M) at (1.5,0) {M};
    \node[node] (N) at (0.5,0) {N};
    \node[node] (O) at (1.5,1)  {O};
    \node[node] (P) at (0.5,1){P};
    \node[node] (Q) at (-0.5,1){Q};

    \draw[edge] (M) -- (O);
    \draw[edge] (M) -- (P);
    \draw[edge] (M) -- (Q);
 \draw[edge] (N) -- (O);
    \draw[edge] (N) -- (P);
    \draw[edge] (N) -- (Q);

    \node[node] (R) at (2,-2) {R};

    \node[node] (S) at (3,-2) {S};
    \node[node] (T) at (2,-3)  {T};
    \node[node] (U) at (3,-3){U};
    \node[node] (V) at (4,-3){V};

    \draw[edge] (R) -- (T);
    \draw[edge] (R) -- (U);
    \draw[edge] (R) -- (V);
 \draw[edge] (S) -- (T);
    \draw[edge] (S) -- (U);
    \draw[edge] (S) -- (V);

    \node[node] (W) at (-1,-2) {W};
    \node[node] (X) at (-2,-2) {X};
    \node[node] (Y) at (-1,-3)  {Y};
    \node[node] (Z) at (-2,-3){Z};
    \node[node] (L) at (-3,-3){L};

    \draw[edge] (W) -- (Y);
    \draw[edge] (W) -- (Z);
    \draw[edge] (W) -- (L);
 \draw[edge] (X) -- (Y);
    \draw[edge] (X) -- (Z);
    \draw[edge] (X) -- (L);

    \draw[edge, color=red] (W) -- (F);
    \draw[edge, color=red] (W) -- (M);
    \draw[edge, color=red] (W) -- (A);
    \draw[edge, color=red] (R) -- (F);
    \draw[edge, color=red] (R) -- (M);
    \draw[edge, color=red] (R) -- (A);
    
\end{tikzpicture}
    \caption{An illustration of the hierarchical construction of a graph with $n=25$ nodes based on $k=5$ graphs of an efficient family  with 5 nodes and 6 edges (Family 10).}
    \label{fig:graph_hierarchical_example}
\end{figure}

Fig.~\ref{fig:graph_hierarchical_example} illustrates a construction of a graph with $n=25$ nodes based on five graphs of 5 nodes, each in the form of  Family 10 in Fig.~\ref{fig:visual_illustration_of_families}. Recall that  this family had low values of the cost functions (Table~\ref{tab:families_cost_ranking}). 

We deduce interesting properties also for Construction II:
\begin{property}
Consider Construction II. 

(i) The diameter is at most $3 \cdot D$, where $D=2$ is the diameter of Family 10. 

(ii) The construction allows establishing pairs of link-disjoint paths also between all pairs of nodes from two distinct copies of the $n$-node graph. 

(iii) In the new construction, all nodes besides those which are connecting the families (A, F, M, R, W) are non-critical nodes.
\end{property}

The proof is similar to the proof of the corresponding property for Construction I with the following changes: As there is no direct edge between all copies of Family 10, the diameter is larger. This also makes one node in each copy of the family to be a critical-node. 

Overall, Construction II allows connecting a similar number of nodes but with fewer edges compared to Construction I. However, this saving in the number of nodes, results in a small increase in the diameter and implies that some of the nodes be critical nodes. 

\textbf{High-level Evaluation of Simple  Datacenter Topologies.} 
We suggest benchmarking against several baseline topologies that are realistic in a datacenter setting. We focus on Construction I rather than Construction II for comparing topologies with almost identical number of nodes. 

We make the following comparison: 
(a) Graph from Fig.~\ref{fig:graph_merge_example} (with red edges) based on Construction I, 

(b) small fat-tree~\cite{Al-FaresLV08} (16 nodes in four groups of 4 each, connected to 4 rail nodes and 2 upper spines), 

(c) Dragonfly+~\cite{DragonflyPlus} (with four groups of 4 leaves), 

(d) Torus~\cite{google_tpu} (of size $2 \times 2 \times 4$) and 

(e) ring. 

For each topology, we can report in Table~\ref{tab:datacenter_comparison} graph metrics (like Table~\ref{tab:n_5_families_basics}) that shows the efficiency of Construction I. 
\begin{table}[h!]
\centering
\scriptsize
\resizebox{0.45\textwidth}{!}{%
\begin{tabular}{c|ccccc}
\hline
Topology & Fig. 7 & Fat-tree & Dragonfly+ & Torus &  ring \\
\hline
\# nodes & 15 &  16 & 16 & 16  & 16  \\
(+\# internal) & -  & 6  & 16 & -  & -  \\
\# links & 24 &  24 & 88 & 32  & 16 \\
\hline
O &  54 & 68 & 152 & 64  & 48\\
R & 1 &  0.812 &  1 &  1  & 1 \\
C & 0.934 & 0.828 & 	1.335 & 0.978  & 0.519\\
\hline
$\Phi$ & 57.82 & 101.14 & 113.86 & 65.44  & 92.49\\
$\Psi$ & 	55.87 & 82.93 & 	131.54 & 	64.72  & 66.67\\
\hline
\end{tabular}
}
\caption{High level comparison of datacenter topologies and simple graph structures for serving QKD. }
\label{tab:datacenter_comparison}
\end{table} 

\section{Conclusions and Future Work} 
This manuscript takes a first step toward topology design for trusted-relay QKD networks by connecting graph structure to key delivery. We propose structural objectives such as operational cost, robustness (non-critical nodes and disjoint paths), and hop distance and then combine them into simple cost functions $\Phi,\Psi$. Using these, we enumerate and rank unlabeled connected graph families for small $n$ (illustrated for $n=5$) and give a simple composition method for larger networks. To tie the graph-theoretic view to physics, we also introduce a lightweight edge model (loss $L_e$ and capacity $K_e \approx \mathrm{SKR}(L_e)$) and highlight the sensitivity of SKR to a few dB of additional loss.

Our future work will employ these tools and move further, even beyond graph theory, by optimizing routing and key management on weighted graphs. Concretely, we plan to (i) compute capacity-aware routing / multi-commodity key flows under heterogeneous $\{K_e\}$, (ii) incorporate explicit trust/risk constraints and multipath key splitting to reduce reliance on any single relay, and (iii) benchmark candidate topologies on realistic datacenter layouts under measured insertion-loss budgets and reconfiguration duty cycles that affect effective key capacity and security.

\bibliographystyle{IEEEtran}
\bibliography{quantum_bib}

\end{document}